\documentclass[useAMS,usenatbib]{mn2e}

\usepackage{graphicx}
\usepackage{natbib}
 
\title [On the discrete nature of the Red Giant Branch of $\omega$~Centauri] {On the
discrete nature of the Red Giant Branch of $\omega$~Centauri
\thanks{Based on FORS observations collected with the Very Large Telescope
at the European Southern Observatory, Cerro Paranal, Chile,  within the
observing program 68.D-0332.}}
\author[A. Sollima , F. R. Ferraro , E. Pancino , M. Bellazzini]{A.
Sollima$^{1,2}$
\thanks{E-mail: antonio.sollima@bo.astro.it},  F. R. Ferraro$^{1}$ , E.
Pancino$^{2}$ , M. Bellazzini$^{2}$\\ $^{1}$ Dipartimento di Astronomia,
Universit\`a di Bologna, via Ranzani 1, Bologna, Italy\\ $^{2}$ Osservatorio
Astronomico di Bologna, via Ranzani 1, I-40127 Bologna, Italy}

\begin{document}

\date{Accepted 2004 January 00. Received 2004 January 00; in original form 2004 January 00}

\pagerange{\pageref{firstpage}--\pageref{lastpage}} \pubyear{2004}

\maketitle

\label{firstpage}

\begin{abstract}
We report the results of an extensive VLT high-resolution imaging campaign of
the central region of the giant globular cluster $\omega$~Centauri 
. More than
100,000 stars have been measured in the inner 9'$\times$9' region of the
cluster. On the basis of multiband Colour Magnitude Diagrams (CMD), we confirm
the existence of multiple stellar populations along the Red Giant Branch (RGB).
Moreover, thanks to the high-precision of this dataset, we conclude
that the RGB does not present a smooth and continuous distribution, but shows a
discrete structure: besides the metal-poor and the extreme metal-rich
population (RGB-a), the existence of three metal intermediate populations is
shown. 
\end{abstract}

 
\section{Introduction}

The understanding of the star formation history of the globular cluster 
$\omega$ Centauri (NGC 5139) represents one of the most interesting challenges
of  stellar astrophysics. $\omega$ Cen is the most massive and luminous GC of
the Milky Way ($M \sim 2.9 ~ 10^6 M_{\sun}$, Merrit et al. 1997) and it is surely
the most peculiar in terms of structure, kinematics and stellar content. One of
the most interesting peculiarities of $\omega$ Cen is its metallicity spread. The
first sign of chemical inhomogeneity was found by Dickens \& Woolley (1967) from
the large colour width of the Red Giant Branch (RGB) in the colour-magnitude
diagram (CMD). This result was later confirmed by a large number of
spectroscopic works (Freeman \& Rodgers 1975, Butler, Dickens \& Epps 1978).

In the last ten years, extensive spectroscopic surveys have been performed on
large samples of giant stars (Norris, Freeman \& Mighell 1996, hereafter NFM96;
Suntzeff \& Kraft 1996, hereafter SK96), showing a multimodal heavy-element
distribution.
Recent wide-field photometric studies (Lee et al. 1999, Pancino
et al. 2000 - hereafter P00, Hilker \& Richtler 2000, Hughes \& Wallerstein 2000,
Rey et al. 2004) have revealed the presence of an  additional, well
defined anomalous RGB (hereafter RGB-a). According to P00, this population
contains approximately 5\% of the cluster's stellar content and  represents the
extreme metal-rich end of the observed metallicity distribution. Moreover, the
metal-rich giants appear to have different spatial distribution and  dynamical
behaviour with respect to the metal poor ones (Pancino et al. 2000, 2003,
Jurcsik 1998, Norris et al. 1997).

In this framework, we have started a long-term project devoted to the detailed
study of the properties of the different stellar populations in this cluster
(see Ferraro, Pancino \& Bellazzini 2002), based on a complete multi-wavelength
photometric survey and an extensive spectroscopic campaign for giant and
subgiant stars, both in the optical and in the infrared bands. Within this
project, a number of results have been published on the photometric, chemical
and kinematic properties of the RGB-a stars (see Pancino et al. 2000, 2002,
2003; Ferraro et al. 2002; Sollima et al. 2004; Origlia et al. 2002,
2003) and the possible detection of  its Sub Giant
Branch (SGB, Ferraro et al. 2004). 

In this paper we present BVI photometry of more than 100,000 stars in the
central region of the cluster \footnote{The entire catalog is only available in 
electronic form at the CDS via anonymous ftp to cdsarc.u-strasbg.fr (130.79.128.5)
or via http://cdsweb.u-strasbg.fr/}.  In \S 2 we describe the observations, the data
reduction techniques and the photometric calibration. In \S 3 we present the
CMD, describing the main characteristics of the different cluster populations.
In \S 4 the photometric metallicity distribution for a large sample of red
giants is presented and discussed.  In \S 5 we show the detection of the RGB
bumps  of the different populations of $\omega$ Cen. Finally, we summarize
and discuss our results in \S 6.

\section{Observations and data reduction}

\begin{figure}
\includegraphics[width=8.7cm]{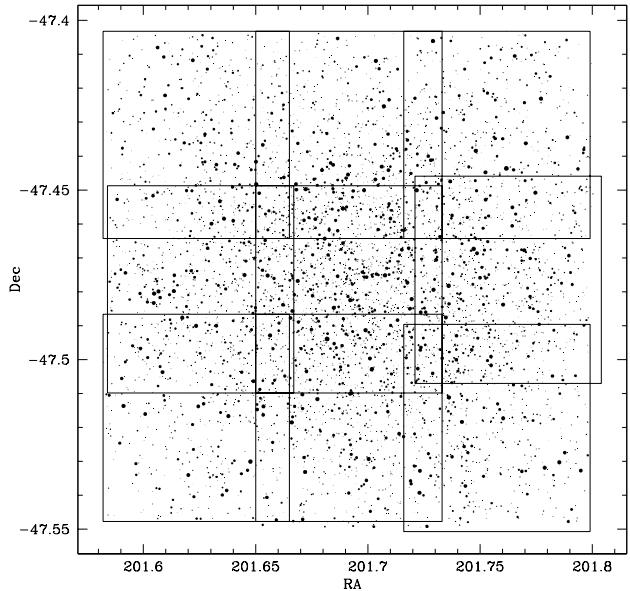}
\caption{Map of the region sampled by the observations. North is up, east on
the right. Only stars with $V<16$  have been plotted.} 
\label{map}
\end{figure}

The photometric data were obtained with the FORS1 camera, mounted at the Unit1
(UT1) of the Very Large Telescope (VLT) of the ESO (Cerro Paranal, Chile). All
observations were performed in Service Mode in 5 different nights on March,
April and June 2002 (see Table 1), using the high resolution mode of FORS1.
In this configuration the image scale is 0.1" $pixel^{-1}$ and the camera has a
global field of view of $3.4' \times 3.4'$ . A mosaic of 9 ($3 \times 3$)
partially overlapping fields has been observed around the cluster center,
covering a total area of $\sim 9' \times 9'$, (see Figure \ref{map}).   

After applying the standard bias and flat-field correction, we used DAOPHOT II
and the point-spread-function (PSF) fitting algorithm ALLSTAR (Stetson, 1987)
to obtain instrumental magnitudes for all the stars detected in each frame. 
For each field, six different frames have been observed (4 long and 2
short-exposure) through the B, V and I filters. Repeated exposures have been
combined in order to obtain a high signal-to-noise median image. The automatic
detection of sources was performed on the median frame adopting a 3$\sigma$
treshold. The mask with the object positions was then used as input for the
PSF-fitting procedure,  that was performed independently on each image. As
usual, the most isolated and bright stars in each field have been employed to
build the PSF model (here a Moffat function of exponent 2.5). For each
passband, the magnitudes obtained from the long exposures were transformed to
the same instrumental scale and averaged. The same procedure was followed for
the short exposures, that were however used only to measure bright stars, which
are saturated in the long exposures. The calibrating equations linking the
instrumental magnitudes to the standard system were obtained from the
comparison with the photometric catalog by Pancino et al. (2000, 2003).
Finally, a catalog with more than 100,000 calibrated stars in the cluster has
been produced. 
Our calibration has been compared with the photometric catalog by Rey et
al. (2004) in the B and V passbands. The mean magnitude differences found are
$\Delta$B = 0.069 $\pm$ 0.066 and $\Delta$V = 0.060 $\pm$ 0.054 , which 
are consistent with a small systematic offset in both passbands.

\begin{center}
\begin{table}
 \centering
 \begin{minipage}{140mm}
  \caption{ Observing logs}
  \label{obs_log}
  \begin{tabular}{@{}llccr@{}}
  \hline
   Field & Date & Filter & Exp time & Seeing (FWHM)\\ 
       & & & (sec) & (")\\
   
 \hline
1 & \rm{14-15~Jun~2002} & B & 299  & 0.6 \\
1 & \rm{14-15~Jun~2002} & B & 10  & 0.6 \\
1 & \rm{14-15~Jun~2002} & V & 45  & 0.6 \\
1 & \rm{14-15~Jun~2002} & V & 2  & 0.6 \\
1 & \rm{14-15~Jun~2002} & I & 25  & 0.6 \\ 
1 & \rm{14-15~Jun~2002} & I & 1  & 0.6 \\ 
2 & \rm{13-14~Jun~2002} & B & 299  & 0.6 \\
2 & \rm{13-14~Jun~2002} & B & 10  & 0.6 \\
2 & \rm{13-14~Jun~2002} & V & 45  & 0.6 \\
2 & \rm{13-14~Jun~2002} & V & 2  & 0.6 \\
2 & \rm{21~Apr~2002} & I & 25  & 0.8 \\
2 & \rm{21~Apr~2002} & I & 1  & 0.8 \\
3 & \rm{14-15~Jun~2002} & B & 299  & 0.6 \\
3 & \rm{14-15~Jun~2002} & B & 10  & 0.6 \\
3 & \rm{14-15~Jun~2002} & V & 45  & 0.6 \\
3 & \rm{14-15~Jun~2002} & V & 2  & 0.6 \\
3 & \rm{14-15~Jun~2002} & I & 25  & 0.6 \\ 
3 & \rm{14-15~Jun~2002} & I & 1  & 0.6 \\ 
4 & \rm{13-14~Jun~2002} & B & 299  & 0.6 \\
4 & \rm{13-14~Jun~2002} & B & 10  & 0.6 \\
4 & \rm{13~Jun~2002} & V & 45  & 0.6 \\
4 & \rm{13~Jun~2002} & V & 2  & 0.6 \\
4 & \rm{20~Apr~2002} & I & 25  & 0.6 \\
4 & \rm{20~Apr~2002} & I & 1  & 0.6 \\
5 & \rm{11~Mar~2002} & B & 299 & 0.6 \\
5 & \rm{11~Mar~2002} & B & 10  & 0.6 \\
5 & \rm{11~Mar~2002} & V & 45  & 0.6 \\
5 & \rm{11~Mar~2002} & V & 2  & 0.6 \\
6 & \rm{11~Mar~2002} & B & 299 & 0.6 \\
6 & \rm{11~Mar~2002} & B & 10  & 0.6 \\
6 & \rm{11~Mar~2002} & V & 45  & 0.6 \\
6 & \rm{11~Mar~2002} & V & 2  & 0.6 \\ 
6 & \rm{11~Mar~2002} & I & 25  & 0.6 \\
6 & \rm{11~Mar~2002} & I & 1  & 0.6 \\
7 & \rm{14-15~Jun~2002} & B & 299  & 0.6 \\
7 & \rm{14-15~Jun~2002} & B & 10  & 0.6 \\
7 & \rm{14-15~Jun~2002} & V & 45  & 0.6 \\
7 & \rm{14-15~Jun~2002} & V & 2  & 0.6 \\
7 & \rm{14-15~Jun~2002} & I & 25  & 0.6 \\ 
7 & \rm{14-15~Jun~2002} & I & 1  & 0.6 \\ 
8 & \rm{11~Mar~2002} & B & 299 & 0.6 \\
8 & \rm{11~Mar~2002} & B & 10  & 0.6 \\
8 & \rm{11~Mar~2002} & V & 45  & 0.6 \\
8 & \rm{11~Mar~2002} & V & 2  & 0.6 \\ 
8 & \rm{11~Mar~2002} & I & 25  & 0.6 \\
8 & \rm{11~Mar~2002} & I & 1  & 0.6 \\
9 & \rm{13-14~Jun~2002} & B & 299  & 0.6 \\
9 & \rm{13-14~Jun~2002} & B & 10  & 0.6 \\
9 & \rm{13-14~Jun~2002} & V & 45  & 0.6 \\
9 & \rm{13-14~Jun~2002} & V & 2  & 0.6 \\
9 & \rm{13-14~Jun~2002} & I & 25  & 0.6 \\
9 & \rm{13-14~Jun~2002} & I & 1  & 0.6 \\

\hline
\end{tabular}
\end{minipage}
\end{table}
\end{center}

\subsection{Photometric errors}

\begin{figure}
\includegraphics[width=8.7cm]{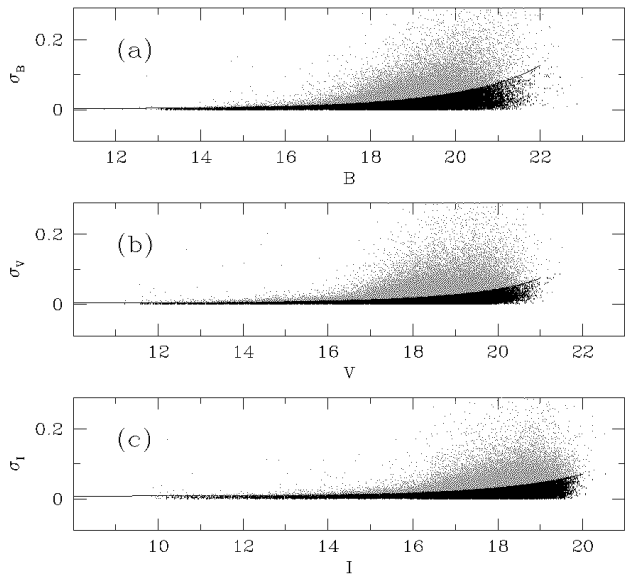}
\caption{B ($panel$ a), V ($panel$ b) and I ($panel$ c) photometric errors as a
function of magnitude. Grey points represent measurements excluded from the
high-accuracy sample.} 
\label{err}
\end{figure}

The root mean square (rms) frame-to-frame scatter of instrumental magnitudes
computed for each star from repeated exposures is a good indicator of the
internal photometric accuracy. Figure \ref{err} reports the distribution of
photometric errors in different passbands as a function of magnitude. As
expected, the errors significantly increase towards faint magnitudes, due to
photon noise. 

We used this diagram in order to define a high-accuracy sample (hereafter HA
sample) that contains only stars with the lowest rms. The behaviour of
photometric errors has been modelled, as a function of magnitude, with
analytical functions, adopting the curves that provide $\sigma_{rms} = 0.01$ at
B=18, V=18 and I=17. To construct the HA sample, only stars with errors below
these curves have been considered. Figure \ref{err} shows the selection
performed on the global catalog: stars included in the HA sample are marked
with heavy black dots.

\section{Color - Magnitude Diagrams}

\begin{figure}
\includegraphics[width=8.7cm]{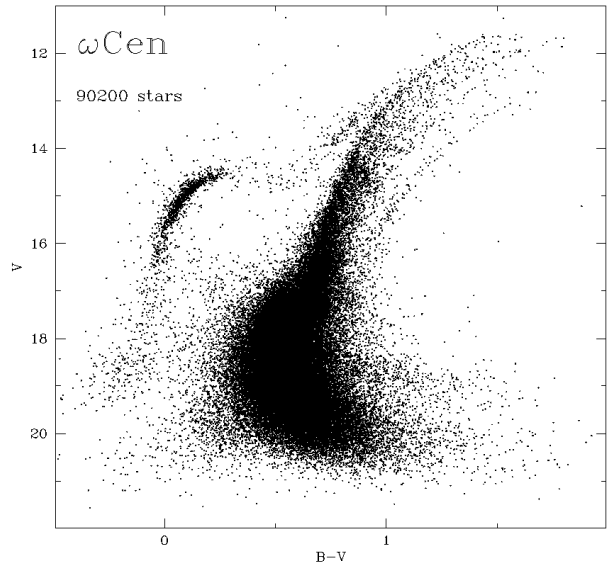}
\caption{$(V,B-V)$ color-magnitude diagram for the global sample of
$\sim$90,000 stars measured in $\omega$ Cen.} 
\label{cmdtot}
\end{figure}

\begin{figure}
\includegraphics[width=8.7cm]{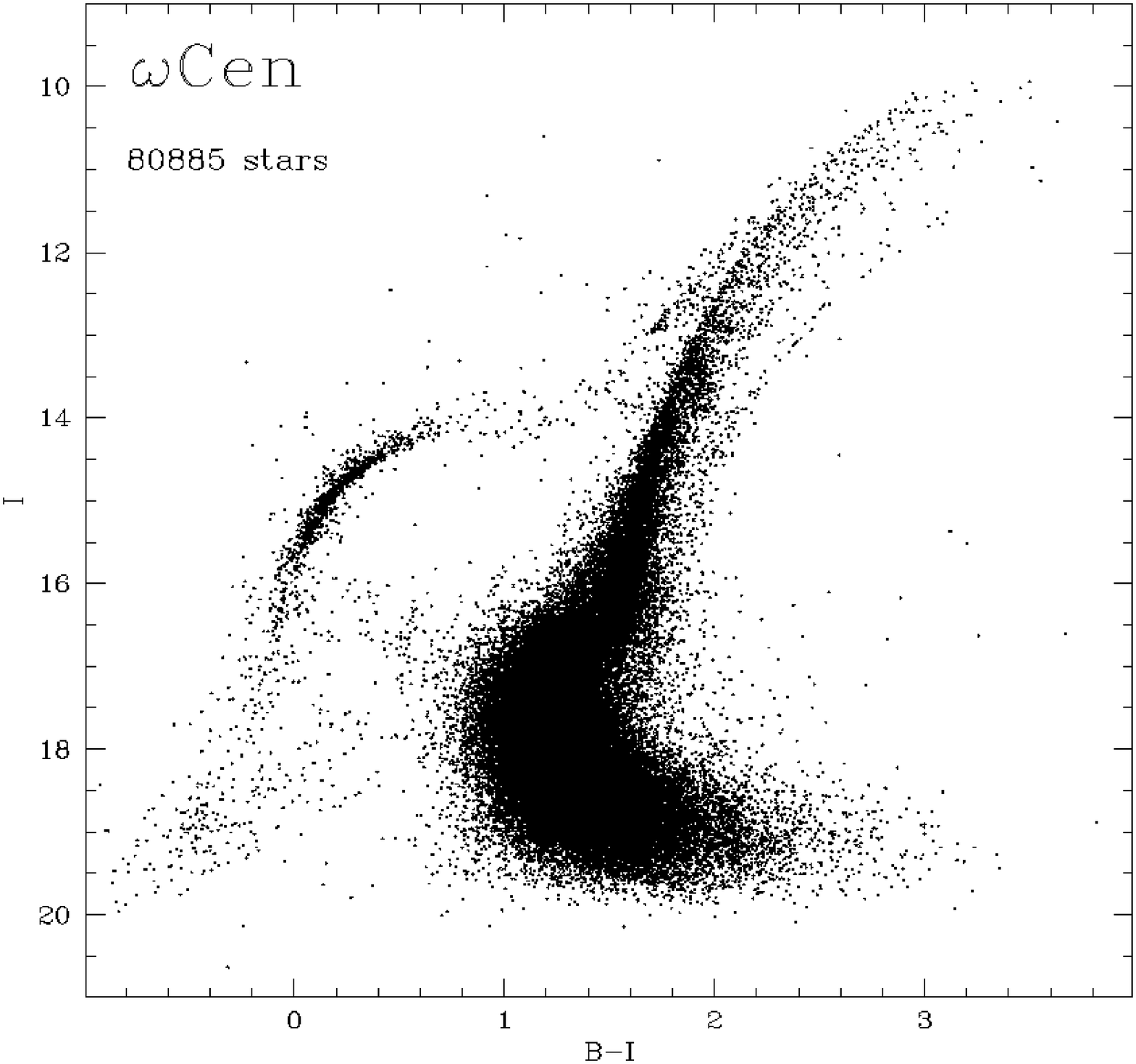}
\caption{$(I,B-I)$ color-magnitude diagram for the global sample of
$\sim$80,000 stars measured in $\omega$ Cen.} 
\label{ibi}
\end{figure}

\begin{figure}
\includegraphics[width=8.7cm]{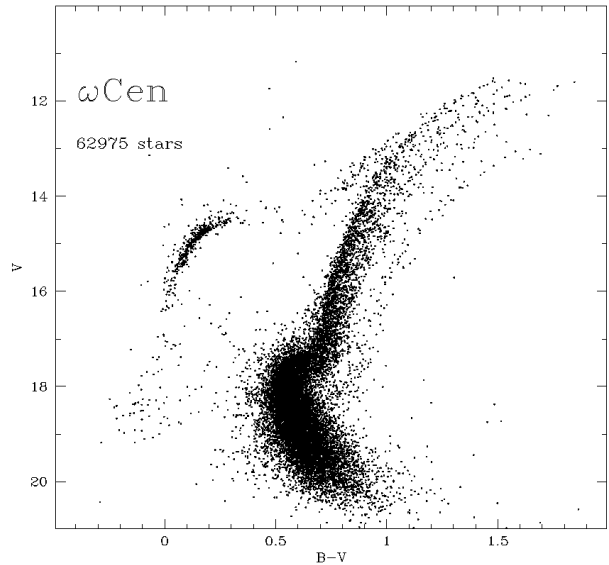}
\caption{$(V,B-V)$ color-magnitude diagram for selcted stars in the
high-accuracy sample (the HA sample).} 
\label{cmd}
\end{figure}

Figures \ref{cmdtot} and \ref{ibi} show the $(V,B-V)$ and $(I,B-I)$ CMDs for
the global sample of stars measured in $\omega$ Cen. Figure \ref{cmd} shows the
$(V,B-V)$ CMD for the high-accuracy sample, where a more precise analysis of
the sequences is possible. The main features of the presented CMDs are
schematically listed below: 

\begin{enumerate}

\item{The CMDs sample the entire evolved population in the cluster, reaching
the upper Main Sequence (MS) at $V\sim 20$;} 

\item{The anomalous RGB (RGB-a, as defined by P00), is also visible as a well
separated population at the extreme red side of the main RGB structure; }

\item{A complex structure of the intermediate RGB components (RGB-MInt(s)),
which  populate the CMD region between the RGB-MP and the RGB-a, can also be
noticed;}

\item{A narrow, well-defined Sub Giant Branch which merges into the MS of the
dominant cluster population is visible in  Figure \ref{cmd}, at a magnitude
level significantly fainter than the MS Turn-Off (this feature is described and discussed
in Ferraro et al. 2004).} 

\end{enumerate}

Apart from the anomalous SGB, the most surprising feature of the CMDs shown in 
Figures \ref{cmdtot}, \ref{ibi} and \ref{cmd} is the {\em discrete structure}
of the intermediate RGBs. The complex morphology of the RGB has been already
noted in several recent publications (e.g. Lee et al. 1999, P00 and Rey et al.
2004). However, in all the above studies, the RGB-MInt population appears to
uniformly populate the CMD region delineated by the dominant RGB-MP population
and the recently discovered RGB-a.

\begin{figure}
\includegraphics[width=8.7cm]{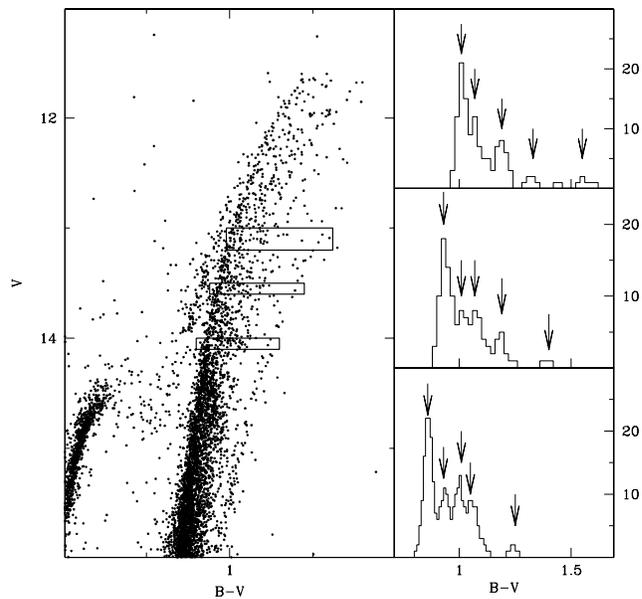}
\caption{RGB colour distribution at different magnitude levels ($V=13,13.5$
and $14$ respectively), as marked on the left $panel$. 5 different peaks are 
evident in the right $panels$, marked by arrows.} 
\label{rgbstruc}
\end{figure}

In the CMDs presented here instead a discrete, comb-like structure becomes evident,
with a set of different intermediate RGB-MInt(s) populating the region between
the RGB-MP and the RGB-a. Figure \ref{rgbstruc} shows a selection of RGB stars,
done on the high-accuracy sample, at three different magnitude levels
($V=13,13.5$ and $14$ respectively). The corresponding colour distributions are
shown in three adjacent panels. As can be seen, five peaks can be identified:
the two extreme peaks being the RGB-MP and the RGB-a, while three additional
peaks can be distinguished between them (RGB-MInt(s)).

This discrete structure of the RGB has strong implications for the chemical evolution of
$\omega$~Cen, and suggests a star formation history characterized by 
bursts, as will be discussed in \S 7. 
  
\section{Metallicity Distribution Function}

\begin{figure}
\includegraphics[width=8.7cm]{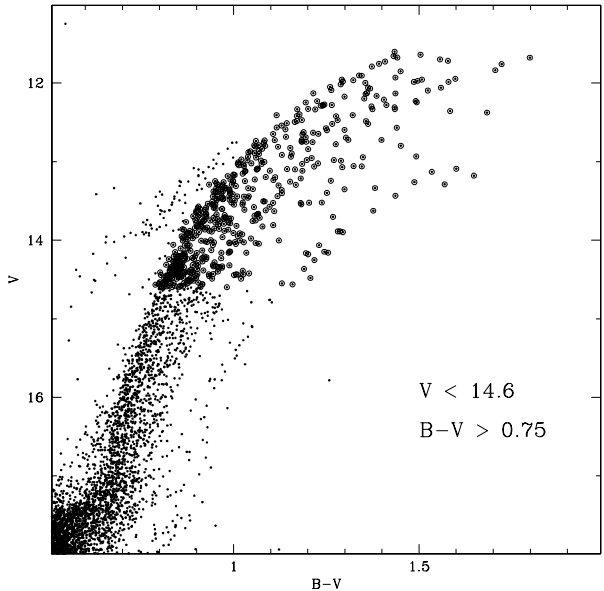}
\caption{The zoomed part of the high-accuracy ($V,B-V$) CMD in the RGB region.
Stars selected to derive the MDF are marked with open circles.}
\label{selbox}
\end{figure}

It is well known that the RGB colour of an old Simple Stellar Population (SSP)
is mainly affected by the abundance of heavy elements and only to a lesser
extent by the age of  the population. The effect of age differences becomes
smaller and smaller with increasing age, and for population II stars it can be
neglected. Therefore, we used the RGB colour distribution to derive a
metallicity distribution function (MDF -- see also Bellazzini et al, 2003). For
this purpose, we selected a sub-sample of stars from the HA sample in the
($V,B-V$) plane, having $V<14.6,~B-V>0.75$ (see Figure \ref{selbox}), in order
to limit the analysis to the region in which colour is most sensitive to
metallicity. We took special care in excluding HB and AGB stars, which are
easily identified from the high quality CMDs shown in Figures \ref{cmdtot},
\ref{ibi}  and \ref{cmd}. 

We then compared the selected RGB stars with a grid of ridge lines of galactic
globular cluster (GGC), having  different metallicities, taken from the sample
of Ferraro et al. (1999 -- hereafter F99). A metallicity estimate was derived
for each star by interpolating the grid of templates in the ($V,B-V$) plane. A
similar approach was adopted by Frinchaboy et al. (2002, hereafter Fr02) who
compared the RGB star distribution from the Washington $M, T_2$ photometry with
theoretical isometallicity curves. Also Lee et al. (1999) and P00 obtained
empirical MDFs in $\omega$ Cen by calibrating, in terms of  metallicity, the
distance of each giant from the mean ridge line of the main metal-poor
component (RGB-MP). Finally, Hilker \& Richtler (2000) derived the photometric
metallicity for a large sample of stars using $vby$ Str\"omgren photometry and
the related metallicity indices. In the following we analyze in detail the obtained
MDF and compare our results with those of the authors quoted above and with the
existing spectroscopic works of NFM96 and SK96. 

\subsection{Basic assumptions}

In order to compare the RGB stars distribution with the mean ridge lines from
F99, we need to assume a  distance modulus and a reddening correction. In the
following we adopt the distance modulus
by Thompson et al. (2001) $(m-M)_0 =13.65 \pm 0.11$ based on the observations
of the detached eclipsing binary system  OGLE~GC17. Concerning the reddening
and extinction coefficients, we assumed $E(B-V)=0.11 \pm 0.01$ (Lub, 2002),
$A_B=4.1E(B-V)$ and $A_V=3.1E(B-V)$ (Savage \& Mathis, 1979).

In our analysis, we considered the global metallicity scale adopted by F99.
Straniero \& Chieffi (1991) and Salaris,
Chieffi \& Straniero (1993) showed that, when computing the isochrones of
Population II stars, the contribution of the $\alpha$-element enhancement can
be taken into account by simply rescaling standard models to the global
metallicity [M/H], according to the following relation 
\begin{equation}
 \mathrm{[M/H]}~=~\mathrm{[Fe/H]}+\log_{10}(0.638~10^{[\alpha/Fe]}+0.362)
\end{equation} 
Direct measurements of the $\alpha$-elements abundance in halo and disk field
stars have shown a very well defined behavior as a function of [Fe/H], with a
nearly constant  overabundance ($[\alpha /Fe] \sim 0.3$) for $[Fe/H] <-1$, and
an evident trend with metallicity, which linearly  decreases to $[\alpha /Fe]
\sim 0.0$ as metallicity increases (see Edvardsson et al. 1993, Nissen et al.
1994,  Magain 1989, Zhao \& Magain 1990, Gratton et al. 1996). Since there is
evidence (Pancino et al. 2002, Origlia et al. 2003) that the stellar
populations in $\omega$~Cen have different $\alpha$-element enhancement levels,
the global metallicity $[M/H]$ is the most appropriate quantity to consider in 
the present case.

\subsection{Results}

\begin{figure}
\includegraphics[width=8.7cm]{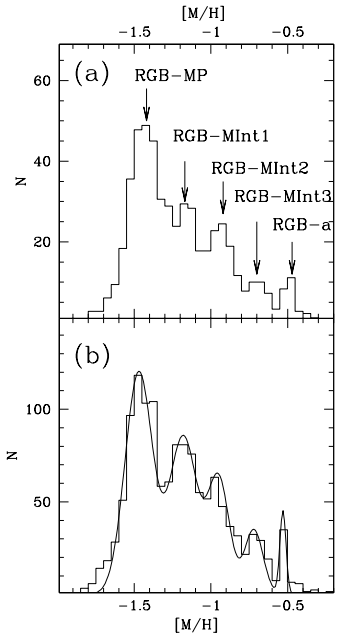}
\caption{ The derived MDF for the high-accuracy (a) and for the global sample
(b). The significant peaks of the distribution are indicated in $panel$ a and
interpolated with a multi-burst enrichment model ($panel$ b).} 
\label{mdf}
\end{figure}

Figure \ref{mdf} shows the obtained MDF for 540 stars in the HA sample ($panel$
a) and for 1,364 stars in the global sample ($panel$ b), with V$<$14.6,
B--V$>$0.75. The distributions are clearly asymmetric: beside the main peak, in
both cases other four peaks are well visible. They can be associated with the
following components:

\begin{enumerate}

  \item{The dominant metal poor population RGB-MP with $[M/H] \simeq -1.4$;}

  \item{Three different metal intermediate components at $[M/H] \simeq -1.2$
  (RGB-Mint1), $[M/H] \simeq -0.9$ (RGB-Mint2) and $[M/H] \simeq -0.7$
  (RGB-Mint3), respectively;}

  \item{The anomalous population RGB-a is clearly visible as a sharp peak at
  the extreme of the distribution (at $[M/H] \simeq -0.5$).}

\end{enumerate}

In order to check the validity of this approach, we compared our metallicity
estimates with those obtained with the Ca Triplet survey by NFM96 (with the 41
stars in common). To convert the metallicities presented by NFM96 into the
global scale, we first converted the equivalent widths of NFM96 into the
Rutledge et al. (1997) system, and derived the iron abundance in the
Carretta \& Gratton (1997) scale (see Section 5 in Rutledge et al. 1997). 

\begin{figure}
\includegraphics[width=8.7cm]{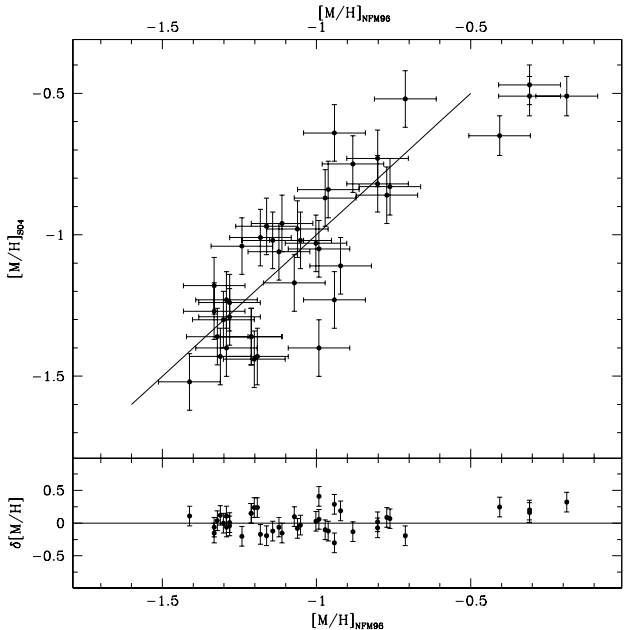}
\caption{Comparison between the metallicities derived in this paper with those
obtained by NFM96 for the 41 common stars (top panel). The residuals of the
comparison are also shown (bottom panel). A solid line in both panels shows the
iso-metallicity locus as a reference.} 
\label{spec}
\end{figure}

Then the derived metallicities have been converted into the global metallicity
scale [M/H] using the eq. (1). According to Ferraro et al. (1999), we assumed
$[\alpha/Fe]=+0.28$ for stars with $[Fe/H]_{CG}<-0.8$ and $[\alpha/Fe]=-0.35
[Fe/H]_{CG}$ for stars with $[Fe/H]_{CG}>-0.8$ (see F99). The comparison of
NFM96 metallicities transformed adopting this procedure with our photometric
metallicities is shown in Figure \ref{spec}. As can be seen, the agreement is
quite good. The solid line shown in Figure \ref{spec} represents the
iso-metallicity relation. The bottom panel shows the residuals of the
comparison. The nice agreement between the two metallicity determinations is
confirmed within the errors, with a mean dispersion of $\sigma_{[M/H]} \simeq
0.2$ dex.

The MDF shown in Figure \ref{mdf} (a) is in good agreement with those presented
by the other photometric surveys (Lee et al. 1999, P00 and Fr02). The peak
detected at $[M/H] \simeq -0.5$ was not found by Fr02, probabily because, as
pointed out by the author, their catalog is severely incomplete in the inner 6
arcmin, a region where the anomalous metal rich RGB-a seems to be concentrated
(see P00 and Pancino et al. 2003, hereafter P03). With respect to the
spectroscopic works (NFM96 and SK96) and the photometric estimates by Hilker
\& Richtler (2000) we note that, while the global shape of the MDFs are similar,
the metal rich components of the distribution presented in this paper are not
reproduced. The lack of metal rich stars in the distributions of NFM96 and SK96
is probably due to two different effects: (i) the different mean radial distances 
of the samples: NFM96 and SK96 limit their analysis to outer regions of the
cluster where the presence of RGB-a stars is less evident. 
(ii) the magnitude selection adopted by NFM96 ($V<13$), which tends to exclude 
RGB-a stars. Note that a few RGB-a stars are visible in the extreme red portion
of the CMD selection boxes of SK96. 
Hilker \& Richtler (2000) found a distribution similar to that of NFM96, with a
continuous high-metallicity tail at $[Fe/H]>-0.8$. The different structure
found at high-metallicities is probably due to the dependence of the
Str\"omgren photometric metallicity index on CN abundance, which is not
linearly dependent on metallicity.
    
\subsection{Statistical significance of the peaks}

As quoted above, the MDF shown in Figure \ref{mdf} suggests the existence of 5
peaks. Following the procedure described by NFM96, we assumed that the 
chemical evolution of a stellar system can be well represented by a simple model 
of chemical enrichment (Searle \& Sargent 1972, Hartwick 1976, Zinn 1978, NFM96). 
This model considers the evolution of a cloud of gas having initial abundance 
$Z_{0}$, which experiences istantaneous recycling of the ejecta from massive 
supernovae, and from which gas is expelled at a rate proportional to the star 
formation rate. Under these assumption the form of the abundance
distribution is $$f(Z)=\sum_{i=1}^{N}f_{i}(Z)$$ where
\begin{eqnarray}
\nonumber
f_{i}(Z)~&=&~\frac{I_{i}~e^{-(Z-Z_{0i})/(<Z_{i}>-Z_{0i})}}{<Z_{i}>-Z_{0i}}
~~~~\rm{for~} Z>Z_{0i}\\
\nonumber
f_{i}(Z)~&=&~0~~~~~~~~~~~~~~~~~~~~~~~~~~~~~~~~~~~~\rm{elsewhere}
\end{eqnarray}

Where $I_{i}$ represents the intensity of the i-th star formation burst.

In order to estimate the significance of the revealed peaks, we fit the
derived MDF with this model, following the prescription of NFM96.
Assuming $[M/H]=log(Z/Z_{\sun})$, we converted $f(Z)$ into a function of the
global metallicity $f([M/H])$ and smoothed it with a gaussian kernel of
$\sigma$=0.05 (equal to the width of the MDF bin, Zinn 1978). We therefore
performed a maximum likelyhood analysis of the data (see appendix B(c) of
Morrison, Flynn \& Freeman 1991). We treated the fit in terms of three free
parameters for each burst: initial metallicity $Z_{0i}$, average metallicity
$<Z_{i}>$ and burst intensity $I_{i}$. 

The choice of the number N of bursts in the model must be done by balancing
{\em bias} and {\em variance} of the fit: a small number of components leads to
an uncertain fit, while large values of N lead to a more stable fit, that
however might contain unnecessary components. For this purpose, we used the
{\em Bayesian Information Criterion} (BIC -- Schwarz, 1978). Let n (=32) be the
bins of the MDF shown in Fig. \ref{mdf} (a), p (=3N) the free parameters of the
fit and $\ell_{N}$ the log-likelihood function at the maximized parameters
values  $$  \ell_{N} = log ~L_{N} =  \sum_{j=1}^{n} log ~Pr_{j,N} $$
  
Where $Pr_{j,N}$ represents the probability that the number of stars whose
metallicity lies in the j-th bin of the MDF be drawn from a N-bursts
distribution.  Then we choose N to maximize the quantity $$ BIC(N)=
\ell_{N}-\frac{p}{2}~log ~n $$ We observe that BIC(N) reaches its maximum value
when N=4. This implies that the observed peaks in the MDF shown if Fig.~\ref{mdf} (a)
are most probably reproduced by a four component model. 

The above analysis of the peak's significance is entirely based on a simple
model of self enrichment. However, there is observational evidence suggesting
that the case of $\omega$~Cen can be much more complex than this (Norris et al.
1997). In particular, recent results cast some doubts on the hypothesis that
the RGB-a was the direct result of a self-enrichment process within
$\omega$~Cen (Pancino et al. 2002, Ferraro et al. 2002, 2004). In
the following discussion, we will extend our analysis to all of the five components as they
appear from the histograms in Figures~\ref{rgbstruc} and \ref{mdf}, and we
reserve to further address the statistical significance of each of them in the
future. We will also treat and discuss the RGB-a separately, whenever needed.

In order to estimate the relative frequency of each population, we computed the
area covered by each single burst distribution $f(Z_{i})$. Since the selection
performed on the HA sample could have corrupted the relative population ratios,
we use the global sample. The results of the fit are plotted over the
histograms in Figure~\ref{mdf} (b) and summarized in Table~\ref{perc}, where
the peak metallicity and the relative frequency are listed for
each considered population. We note that the extreme metal-rich population
(RGB-a) represents $\sim $5\% of the stars of the whole system, fully in
agreement with P00 and P03. Regarding the metal poor population, its relative
frequency is significantly smaller then the estimate of P03 ($\sim $70\%),
while the Mint(s) populations represent a consistent fraction of the whole
system with respect to what estabilished by P03 ($\sim $25\%). This
effect is most probably due to the contamination of the RGB-MInt1 on the RGB-MP
sample of P03, whose photometry did not allow a clear separation between the
two.

\begin{table}
   \centering
   \begin{minipage}{140mm}
    \caption{Relative frequency of the main populations of $\omega$ Cen}
    \label{perc}
    \begin{tabular}{@{}lcr@{}}
    \hline
      & $[M/H]$ & \% \\
    
   \hline
  RGB-MP & -1.4 & $42 \pm 8$  \\
  RGB-MInt1 & -1.2 &  $28 \pm 6$  \\
  RGB-MInt2 & -0.9 &  $17 \pm 5$  \\
  RGB-MInt3 & -0.7 &  $8 \pm 3$  \\
  RGB-a & -0.5 & $5 \pm 1$  \\
  \hline
  \end{tabular}
  \end{minipage}
\end{table}

\subsection{Spatial distribution}  

\begin{figure}
\includegraphics[width=8.7cm]{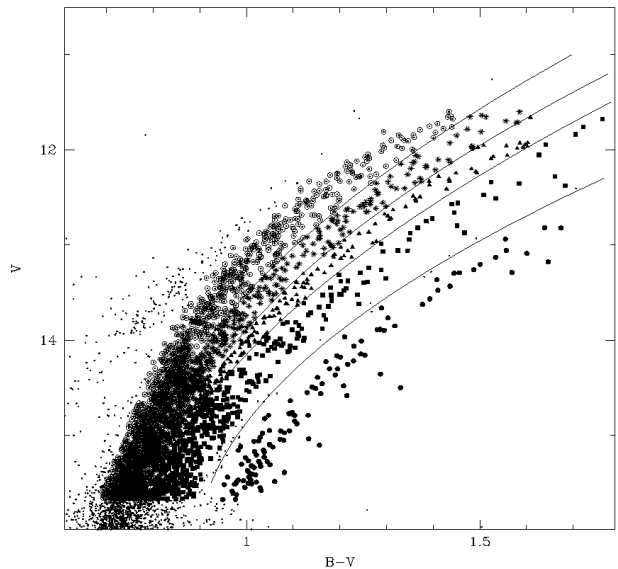}
\caption{Groups selection in the global ($V,B-V$) CMD. The different population
samples are marked with different simbols. The continuous lines follow the
distribution minima and are used to separate the groups.} 
\label{sel}
\end{figure}

Using the RGB samples defined in the previous section, we investigated the
spatial distributions of the sub-populations. In order to minimize the
contamination between the metal-poor and MInts populations we limited our
analysis to the brightest portion of the RGBs ($V<16$, see Fig.~\ref{sel}).
We computed the surface density distributions of each sample using an adaptive kernel 
estimator algorithm (Fukunaga 1972, Silvermann 1986, Beers
1991, Merrit \& Tremblay 1994). However, the smaller area covered by the
photometry presented here, and the subdivision in five components, drastically
reduce the number size of each sample so that only a qualitative discussion can be
done on the resulting spatial distributions. The comparison of the observed
distributions reveals that both the MInts and RGB-a stars present a distribution  
that significantly differs from the RGB-MP sample's one.
A two-dimensional generalization of the Kolmogorov-Smirnov test (Peacock 1983, 
Fasano \& Franceschini 1987) gives a probability that the spatial distribution 
of the co-added (RGB-MInts $+$ RGB-a) sample and the RGB-MP one are drawn from 
the same parent distribution of less than $0.2$ \%. These results confirm those 
of P00, P03 and NFM96.
Table~\ref{center} reports the location of the centroids, the distance from the 
dominant population (RGB-MP) center and the number size of the sample for the 
different population of $\omega$ Cen. We observe that, while the centroid
of the MInt1 sample is located close to the dominant MP component
center, the three most metal rich populations
(RGB-MInt2, RGB-MInt3 and RGB-a) seem to have different centroids. 
In particular, the RGB-a centroid appears displaced $\sim 61"$ to the north of the dominant population, 
in full agreement with P03.

\begin{table}
 \centering
 \begin{minipage}{140mm}
  \caption{ RGB centroids}
  \label{center}
  \begin{tabular}{@{}lcccccr@{}}
  \hline
    & $RA$  & $Dec$ & $\delta$RA & $\delta$Dec & d   & N\\ 
    & (deg) & (deg) &     (deg)    &   (deg)   &(")  &\\
   
 \hline
MP    & 201.703  & -47.4711 & -- & -- & -- & 1132 \\
MInt1 & 201.694  & -47.4865 & -0.009 & -0.0154 & 60 & 637 \\
MInt2 & 201.686  & -47.4850 & -0.017 & -0.0139 & 64 & 411 \\
MInt3 & 201.744  & -47.4669 & 0.041 & 0.0042 & 101 & 244 \\
RGB-a & 201.691  & -47.4551 & 0.012 & 0.0160 & 65 & 133 \\
 
\hline
\end{tabular}
\end{minipage}
\end{table}

\section{The Red Giant Branch Bumps}  

One of the most interesting features of the RGB is the so-called RGB bump
(Iben, 1968). The location of the RGB bump in the CMD is mainly a function of
age and metal content of a given stellar population. Because of the
age-metallicity differences between the populations of $\omega$ Cen, one
expects that this feature will appear in each RGB  at a fainter magnitude as
metallicity increases. The identification of the RGB bump is not an easy task 
because of the need of large observational samples (Crocker \& Rood 1984, Fusi
Pecci et al. 1990). 

\begin{figure}
\includegraphics[width=8.7cm]{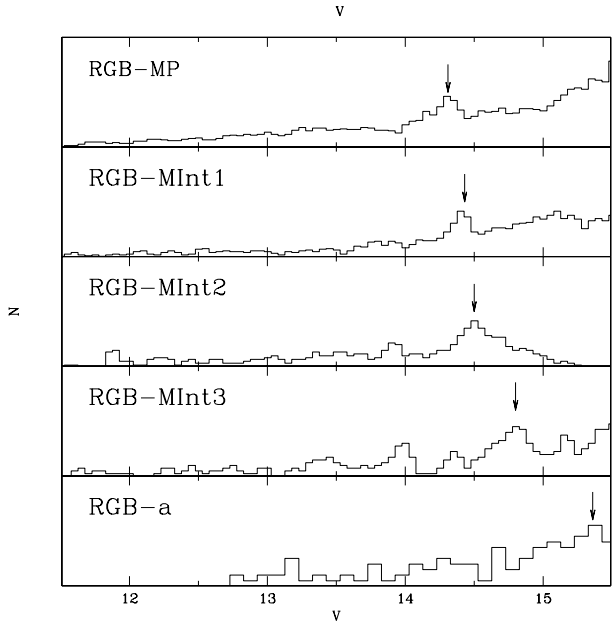}
\caption{Differential LFs for five RGB groups. The peaks corresponding to the
RGB bumps are marked with arrows. In the upper $panel$ the groups selection in
the ($V,B-V$) CMD is shown.} 
\label{bumps}
\end{figure}

To identify the RGB bump in our CMD we performed the procedure described in the
following.  By computing the colour distribution of stars at different
magnitude levels along the RGB in the global ($V,B-V$) CMD, we  separated the
five populations of $\omega$ Cen by following the position of the distribution
minima (see Fig. \ref{sel}). Then we located the RGB bumps by identifying the
corresponding peaks in the differential RGB luminosity  functions (LF).
Figures~\ref{sel} and \ref{bumps} shows the five RGB groups and their
corresponding LFs, respectively. The bumps for the RGB-MP and RGB-Mint 
populations are clearly defined and marked with arrows, while the RGB bump of
the extreme metal rich population (RGB-a) is much less evident and will be not
considered in the following discussion. In Table~\ref{bump_tab} the observed
$V_{Bump}$ magnitudes for the four considered populations of $\omega$ Cen are
listed. 

\begin{figure}
\includegraphics[width=10.5cm]{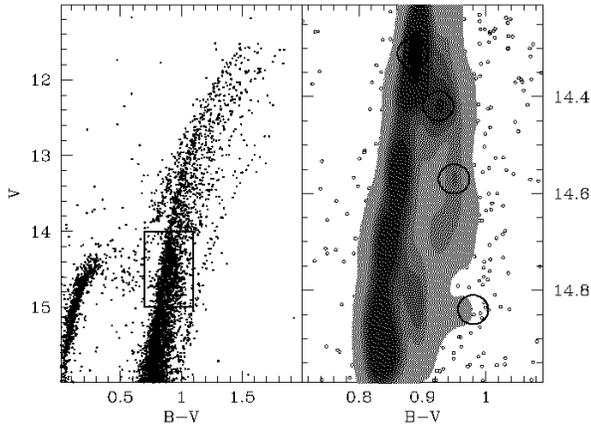}
\caption{Hess diagram for the RGB bump region marked in the left $panel$. In
the right $panel$, the location of the RGB bump for the MP and the three MInts
populations is marked with open circles.} 
\label{hess}
\end{figure}

It is interesting to note that the location of the RGB bumps on our CMD does
not appear as a "continuous and slanting feature", as it was described by Rey
et al., (2004) but as a succession of distinct peaks. This is best put into 
evidence in Figure~\ref{hess}, where the Hess diagram constructed on a zoomed 
area of the ($V,B-V$) CMD is presented. As can be seen, the bumps are clearly 
separated from each other, again pointing towards a set of well separated 
components in the RGB of $\omega$~Cen.

\begin{figure}
\includegraphics[width=8.7cm]{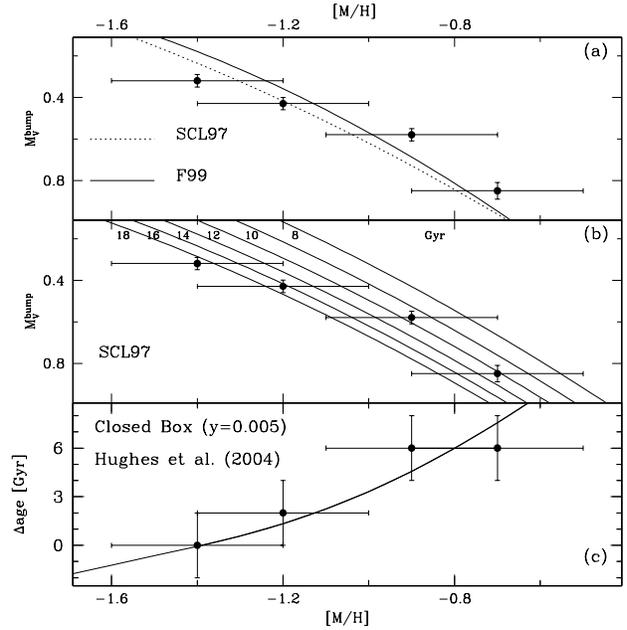}
\caption{ Panel (a): $M_V$ of the RGB bumps as a function of global 
metallicity [M/H]. The solid line is the empirical  relation by F99, 
the dashed line is the theoretical prediction by the Straniero, Chieffi \& 
Limongi (1997) models at t=16 Gyr. Panel (b): Comparison with the isochrones
in the $M_V$ vs $[M/H]$ plane. The isochrones are derived from the same models
that adopt a fixed He abundance of $Y=0.23$ (see F99).
Panel (c): Age-Metallicity Relation obtained from
the bumps. The thick continuous line is the AMR predicted by the same ``closed
box'' model adopted by Hughes et al. (2004), assuming constant star formation
and yeld $y=0.005$.} 
\label{comp}
\end{figure}

The absolute magnitudes of the RGB-bumps are plotted as a function of the
photometric metallicities, obtained in Sect.~4, in panel a of 
Figure~\ref{comp}. The
result is compared to the empirical relation of F99 (obtained from the
detection of the RGB-bump in 54 GGC) and to the theoretical expectation by 
Straniero, Chieffi \& Limongi (1997, hereafter SCL97), computed for $Y=0.23$ 
and age$=16$ Gyr. 
As can be seen (Figure \ref{comp}), the
location of the RGB bump for the four considered populations of $\omega$~Cen is
in agreement, within the uncertainties, with the predictions of both
relations. This further supports the calibration of the derived MDF discussed
in Sect.~4 and the discrete nature of the RGB components in $\omega$~Cen.

In panel (b) of Fig.~\ref{comp} the positions of the bumps are compared with
isochrones in the plane $M_V^{bump}$ vs. $[M/H]$ derived from the SCL97 models
(Eq. 3 of F99). Neglecting the actual zero point of the resulting age-scale,
that may strongly depend on the details of the theoretical models
(see also Riello et al. 2003), the
distribution of the observed points over the isochrones grid suggests an age
range of several Gyr between the MP and the most metal-rich MInt populations.
While still within the observational uncertainties, the observed trend deserves
some discussion, since it provides independent support to the results obtained 
from the analysis of Turn-Off stars with Str\"omgren photometry by Hughes et al. 
(2004), and with medium resolution spectroscopy by Hilker et al. (2004). 

The derived age range ($\sim 6$ Gyr) has to be considered as a strong upper
limit because of (a) the unrealistically old zero-point of the age scale (18 Gyr)
and (b) the neglection of the Helium enrichment which should be associated with
the process that altered the mean metallicity from $[M/H]\simeq -1.4$ to
$[M/H]\simeq -0.7$. 
The more metal rich bumps plotted in
Fig.~\ref{comp} appear brighter than that
expected if they were associated with population having the same age and He
content of the MP population.
Since both a younger age and an increase in He
abundance (Y) should enhance the luminosity of the bumps, the observed trend may
be interpreted either in terms of age and/or Y variations.
In this context, several authors considered the effects of differences in
helium abundance between the different populations of 
$\omega$ Cen (Ferraro et al. 2004, Bedin et al. 2004, D'Antona \& Caloi 
2004, Norris 2004). In particular, Norris (2004) suggested a helium enhancement 
of $\delta Y \sim 0.10 - 0.15$ in order to reproduce the complex MS-TO morphology 
of $\omega$ Cen. SCL97 models predict $$\frac{\delta
M_{V}^{Bump}}{\delta Y} \sim -2.45$$
Assuming that the MInt2 population corresponds to the "second population" of 
Norris (2004) (having $\Delta$[M/H]=0.5 with respect to the dominant MP
population), the observed difference in the RGB bump level between these
populations can be induced by an helium enhancement of $\Delta Y \sim$ 0.09, 
once the effect due to the metallicity has been taken into account and assuming 
no age difference between these populations. This finding is not too far from
what proposed by Norris (2004), but should be considered as an upper limit if we
accept that MInt2 population is younger than MP one, as suggested by elementary
concepts of chemical evolution.  
The most reasonable hypothesis is that both factors are contributing, according
to the natural trend of chemical evolution, e.g. successive generations of stars
are younger than the MP population and more enriched in metals and He.

In panel (c) of Fig.~\ref{comp} the differential Age-Metallicity Relation (AMR)
obtained from panel (b) is displayed ($\Delta$ age $=0.0$ is placed at age$=18$
Gyr). This plot shows that the AMR derived from the RGB bumps is very similar to
that obtained by Hughes et al. (2004) and it is fully compatible with the same 
``closed box'' model adopted by these authors (see their Fig.~7). The above
comparisons strongly indicate that there is full compatibility between the
constraints obtained from the TO region of the CMD by Hughes et al. (2004) and
Hilker et al. (2004). 
and the observed trend in the $M_V^{bump}$ vs. $[M/H]$ described in the present
paper. Both results suggest that the chemical enrichment of the $\omega$
Centauri system in the range $-1.4\le [M/H] \le -0.7$ occurred on a timescale
of a few Gyr ($< 6$ Gyr), the actual value depending on the unknown amount of
Helium enrichment.

\begin{table}
 \centering
 \begin{minipage}{140mm}
  \caption{RGB bump parameters of $\omega$ Cen}
  \label{bump_tab}
  \begin{tabular}{@{}lcr@{}}
  \hline
   RGB bump & $V_{bump}$ & $[M/H]$ \\ 
   
 \hline
MP    & $14.31 \pm0.03$  & $-1.4 \pm0.2$ \\
MInt1 & $14.42 \pm0.03$  & $-1.2 \pm0.2$ \\
MInt2 & $14.57 \pm0.03$  & $-0.9 \pm0.2$ \\
MInt3 & $14.84 \pm0.04$  & $-0.7 \pm0.2$ \\
 
\hline
\end{tabular}
\end{minipage}
\end{table}

\section{Discussion and Conclusions}
 
We presented high precision BVI photometry for $\omega$ Cen, resulting 
from an accurate survey of the cluster's core. More
than 100,000 stars have been measured, allowing a complete characterization
of the RGB. In particular, the most robust conclusion of our work is
that {\em the complex structure of $\omega$~Cen RGB does not show a
smooth and continuous distribution of populations, but it has a
clearly
discrete structure.} At  least four, most probably five, separate
components have been identified. The discreteness and different nature
of the components, clearly visible from the new CMD presented
(Fig.~\ref{selbox}), has been further demonstrated by studying in
detail: {\em (i)} the photometric metallicity distribution function,
which shows five separate peaks (see Fig.~\ref{mdf}) and {\em (ii)}
the identification of the RGB bump for each population (although some
uncertainty remains on the RGB-a bump), which appear as distinct
peaks if studied  through a Hess diagram (Fig.~\ref{hess}), and indicate an age
difference of $\sim$ 6 Gyr between the MP and the most metal-rich MInt population. 

Such a characteristic has one simple implication on the star formation
history of $\omega$~Cen, if interpreted from the point of view of a
simple self-enrichment scenario.
It implies in fact that each
population of $\omega$~Cen (except maybe the RGB-a, see below)
can be associated with a different episode of star formation. Therefore,
the chemical evolution of $\omega$~Cen should be viewed as a series
of bursts, separated in time.
However, a {\em simple} self-enrichment scenario may not be the best
description of the evolution of $\omega$~Cen. We recall at this
point the puzzling correlations between chemical and kinematical properties
found by Freeman (1985) and Norris et al. (1997). Also from the chemical point 
of view $\omega$~Cen shows some peculiarities which are not observed in normally
self-enriching systems, i.e. dwarf galaxies and the field of the Milky Way. One
such characteristic is the presence of CH-CN, Na-O, Mg-Al anomalies (Norris \&
Da Costa 1995a), which up to now has only been oserved in galactic globular
clusters, $\omega$~Centauri and nowhere else. The second is the
strong enhancement of s-process elements seen in both the metal-poor and
intermediate components (Evans 1983, Norris \& Da Costa 1995b, Smith et al.
2000, Vanture et al. 2002) and the RGB-a (Pancino 2003) of $\omega$~Centauri.
This last peculiarity has only been observed in another stellar
system, the Sagittarius dwarf galaxy, which is presently merging with the Milky
Way.
Therefore, a simple self-enrichment cannot explain all the observational 
evidences gathered in the past, in particular when the RGB-a population is
considered.
In fact it appears to stand out from the other populations in terms 
of morphology, structure, chemical composition and kinematics.
The RGB-a main peak appears significantly dislocated
from the main population's one. Moreover, the RGB-a chemical
abundance, besides the higher iron content, presents a smaller
$\alpha$-enhancement with respect to the other populations (Pancino
et al. 2002, P03, Origlia et al. 2003) while a large
overabundance of heavy neutron capture elements remains high as for the other
populations (Pancino 2003). Finally, the mean
proper motion of the RGB-a stars appears significantly different from that of
the main cluster population (Ferraro et al. 2002).
This last result have been questioned by Platais et al. (2003) who claimed
that the different proper motion observed for the RGB-a is due to a
spurious instrumental effect in the original proper motion catalog by van Leuween
et al. (2000). However, while the arguments claimed by Platais et al. (2003)
were already considered and dismissed in the original publication by van Leuween
et al. (2000), they also have been strongly criticized by Hughes et al. (2004) 
who emphasised that no evidence of systematic trend has been detected in the 
proper motion residuals. 
Further support to the correctness of the proper motion measurements has been
brought by Pancino (2003), who again did not found any
significative residual trend as a function of either magnitude or color.

Summarizing, the whole body of evidence points towards a
complex scenario, where the different populations could have not
only had a different chemical enrichment history, but also {\em a
different dynamical evolution}. It is clear that if one
desires to accommodate all the observational facts in one self-consistent
scenario, it is necessary to include not only a complex chemical evolution, but
also a complex dynamical history, possibily including minor mergers  
within a framework such as the "merger within a fragment" (Searle 1977, Norris
et al. 1997), or capture of smaller components. 
 
\section*{Acknowledgments}

This research was supported by the Agenzia Spaziale Italiana and the Ministero
dell'Istruzione, dell'Universit\`a e della Ricerca.
We warmly thank Paolo Montegriffo for assistance during the catalogs
cross-correlation and astrometric calibration procedure and Santino Cassisi and
Oscar Straniero for the helpful scientific discussion.
We also thank John E. Norris, the Referee of our paper, for his 
precious comments and suggestions.

\label{lastpage}

\end{document}